\documentclass[12pt]{article}

\usepackage{graphicx}
\usepackage{graphics}
\usepackage{epsfig}
\usepackage{amssymb}
\usepackage{amsmath}
\usepackage{color}
\usepackage{textcomp}
\usepackage{latexsym}

\usepackage[margin=2.5cm,footskip=0.6cm]{geometry}
\def\bea{\begin{eqnarray}}
\def\eea{\end{eqnarray}}
\def\beq{\begin{equation}}
\def\eeq{\end{equation}}

\def\bm{\begin{math}}
\def\me{\end{math}}

\def\del{\partial}

\begin{document}

\begin{center}
{\Large{\bf Surface-directed Spinodal Decomposition on Morphologically Patterned Substrates}} \\
\ \\
\ \\
by \\
Prasenjit Das$^{1,2}$, Prabhat K. Jaiswal$^3$ and Sanjay Puri$^2$ \\
$^1$Department of Chemical and Biological Physics, Weizmann Institute of Science, Rehovot 76100, Israel. \\
$^2$School of Physical Sciences, Jawaharlal Nehru University, New Delhi 110067, India. \\
$^3$Department of Physics, Indian Institute of Technology Jodhpur, Karwar 342037, India. \\
\end{center}

\begin{abstract}
This paper is the second in a two-part exposition on {\it surface-directed spinodal decomposition} (SDSD), i.e., the interplay of kinetics of wetting and phase separation at a surface which is wetted by one of the components of a binary mixture. In our first paper [P. Das, P.K. Jaiswal and S. Puri, Phys. Rev. E {\bf 102}, 012803 (2020)], we studied SDSD on chemically heterogeneous and physically flat substrates. In this paper, we study SDSD on a chemically homogeneous but morphologically patterned substrate. Such substrates arise in a vast variety of technological applications. Our goal is to provide a theoretical understanding of SDSD in this context. We present detailed numerical results for domain growth both inside and above the grooves in the substrate. The morphological evolution can be understood in terms of the interference of SDSD waves originating from the different surfaces comprising the substrate.
\end{abstract}

\newpage
\section{Introduction}
\label{sec1}

In the past few decades, the kinetics of phase separation has attracted much research attention. Consider a binary (AB) mixture, which is homogeneous at high temperatures. When this system is quenched below the miscibility gap, it becomes thermodynamically unstable and prefers to be in the phase-separated state. The evolution of the unstable homogeneous mixture proceeds by the emergence and growth of A-rich and B-rich domains \cite{pw09}. For a translationally invariant system, the domain growth morphology is characterized by the {\it order parameter correlation function} $C(\vec{r},t)$, where $\vec{r}$ is the spacing between two points and $t$ is the time. If the system is isotropic and described by a unique length scale $L(t)$, the correlation function exhibits {\it dynamical scaling} \cite{pw09}:
\begin{equation}
\label{scale1}
C(\vec{r},t) = g\left[r/L(t) \right] ,
\end{equation}
where $g(x)$ is the scaling function.

Phase-separating systems are usually studied via scattering experiments, where a suitable probe (e.g., light, X-rays, neutrons, etc.) is scattered. The scattering amplitude is proportional to the {\it structure factor} $S(\vec{k},t)$, which is the Fourier transform of $C(\vec{r},t)$ at wave-vector $\vec{k}$:
\beq
S(\vec{k},t) = \int d\vec{r}~e^{i \vec{k} \cdot \vec{r}}~C(\vec{r},t) .
\eeq
The corresponding dynamical scaling form for $S(\vec{k},t)$ is
\beq
\label{scale2}
S(\vec{k},t) = L^d f\left[kL(t) \right] ,
\eeq
where $d$ is the dimensionality, and $f(p)$ is a scaling function. For diffusion-driven phase separation, the coarsening of domains obeys $L(t) \sim t^{1/3}$, which is known as the {\it Lifshitz-Slyozov} (LS) growth law \cite{ls61,dh86}. In segregating fluids, the material can also be advected via the fluid velocity field. This gives rise to crossovers in the domain growth law as $t^{1/3}$ (diffusive regime) $\rightarrow$ $t$ (viscous hydrodynamic regime) $\rightarrow$ $t^{2/3}$ (inertial hydrodynamic regime) \cite{pw09}.

The above scenario applies to bulk mixtures. However, in experiments, segregating mixtures are always in contact with surfaces. These could be the ``passive walls'' of a container, or ``active walls'' which are designed to preferentially attract one of the components (say, A) of the mixture. The presence of a surface or substrate (denoted as S) can dramatically alter phase-separation kinetics, giving rise to the well-known process of {\it surface-directed spinodal decomposition} (SDSD) \cite{jnk91,gk95,gk03,pf97,kb98,sp05}. This process is of great scientific and technological importance in the context of problems ranging from {\it nano-lithography} to {\it directed self-assembly of polymer mixtures}. Depending upon the strength of surface tensions between A, B and S, the substrate shows either a \textit{partially wet}~(PW) or a \textit{completely wet}~(CW) equilibrium morphology. In the PW morphology, the interface between A-rich and B-rich domains makes a {\it contact angle} $\theta$ with the substrate S, which is determined by {\it Young's condition} \cite{ty05}:
\begin{eqnarray}
\label{young}
\sigma \cos\theta = \gamma_{\rm BS} - \gamma_{\rm AS},
\end{eqnarray}
where $\sigma$, $\gamma_{\rm AS}$ and $\gamma_{\rm BS}$ are the A-B, A-S and B-S surface tensions, respectively. In the CW morphology, Eq.~(\ref{young}) does not have a solution ($\gamma_{\rm BS} - \gamma_{\rm AS} > \sigma$), and the AB interface becomes parallel to the substrate.

The problem of SDSD on chemically homogeneous and physically flat substrates has been studied extensively via experiments~\cite{jnk91,gk95,gk03} and simulations \cite{pb92,pbf97,pb01,be90,bc92,jm93,fmd97,fmd98}. The first successful coarse-grained model for SDSD was proposed by Puri and Binder (PB) \cite{pb92}, who supplemented the Cahn-Hilliard-Cook (CHC) model of phase separation with two boundary conditions which modeled the surface. PB showed that the surface gives rise to an SDSD wave, in accordance with the experiments of Jones et al. \cite{jnk91}. The SDSD wave consists of alternating wetting and depletion layers of the preferred component, and propagates into the system. PB focused on two important aspects of the SDSD morphology: \\
(a) the growth law for the wetting layer thickness $R_1(t)$; \\
(b) the layer-wise correlation functions $C(\vec{\rho},z,t)$, and structure factors $S(\vec{k_\rho},z,t)$. Here, we have decomposed coordinates as $\vec{r} = (\vec{\rho}, z)$, where $\vec{\rho}$ and $z$ are the coordinates parallel and perpendicular to the surface (located at $z=0$).

PB showed that $R_1(t)$ has an early-time behavior, which depends on the surface potential $V(z)$ \cite{pb01}. In the asymptotic regime, $R_1(t)$ shows a crossover to the universal LS behavior. PB also showed that, for $z$-values outside the wetting/depletion layers, the layer-wise correlation functions and structure factors exhibit dynamical scaling. The layer-wise domain size $L(z,t)$ follows the LS law, but the prefactor is higher near the surface. PB explained this as a result of the alignment effects of the layered SDSD profile at the surface.

In recent work, we have initiated a theoretical study of SDSD on chemically and morphologically patterned substrates. Let us first briefly discuss the case of {\it chemically patterned surfaces}, which we addressed in a recent paper \cite{djp20}. There are many experimental studies of this problem, which has myriad technological applications. Some of these have been cited in Ref.~\cite{djp20}. There have also been some theoretical (simulation) studies of this problem. However, to our surprise, most of the theoretical studies focused on a pictorial classification of the morphologies which arise on chemically patterned surfaces. In Ref.~\cite{djp20}, we performed a quantitative study of the emergent morphologies in terms of growth laws and scaling functions. Most of our results were obtained in the context of a checkerboard substrate, but were generic to a wide range of surface patterns.

This paper is the second in a two-part exposition. Here, we focus on the case of {\it morphologically patterned} or {\it rough surfaces}, i.e., the surface is chemically homogeneous but has a non-uniform spatial structure. This roughness could either arise naturally or be designed to serve a specific functionality. An important application of morphologically patterned surfaces is the transport of fluid mixtures through narrow channels at micron length scales \cite{kyb02,kb03}. These {\it microfluidic systems} lie at the heart of the lab-on-a-chip concept, which promises to integrate all the functions of a conventional chemistry laboratory into a single compact chip \cite{rs98,kvb98,mg99,ws01,rim02,mh02,gw06,hs09}. These devices find application in diagnostic testing, forensics, DNA manipulation, food industries, micro-reactors, etc. The emergence of {\it micro-contact printing} has facilitated the fabrication of such devices. This technique can be employed in designing a physical pattern on a surface, which consists of posts of different shapes and sizes arranged in a required form~\cite{kbw94,jww95}. There also exist important biological applications of microfluidic networks \cite{dbs97,kto99}. Apart from microfluidic systems, morphologically patterned surfaces are also very useful in quantum dots, nanowires, magnetic storage media, nanopores and silicon capacitors \cite{kss03}. In a related context, Hochrein et al. \cite{hlg06} have reported interesting experiments on DNA stretching (a form of coarsening) on a 1-dimensional rippled surface.

\begin{figure}[t]
\centering
\includegraphics[width=0.40\textwidth]{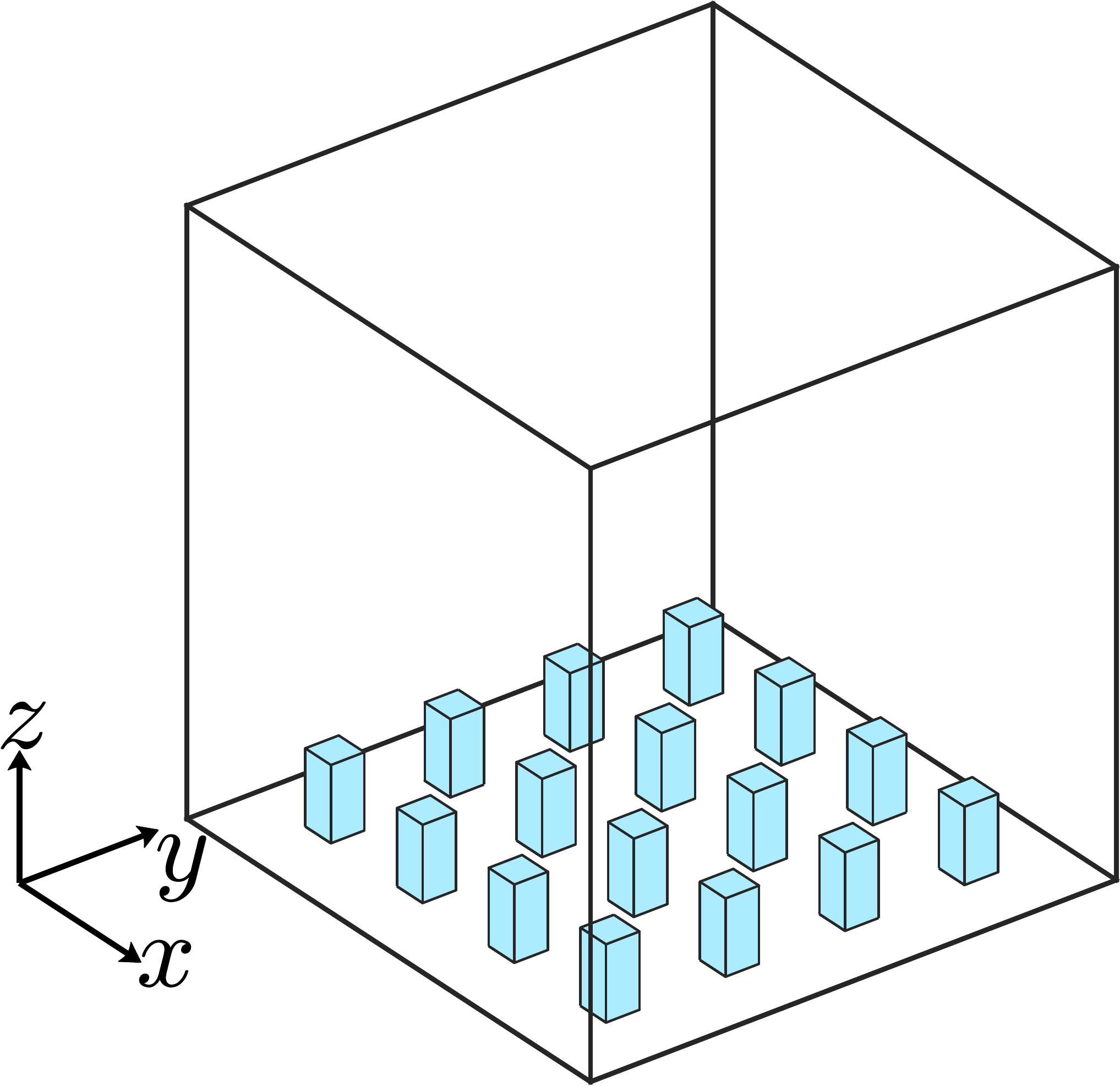}
\caption{\label{fig1} (color online) Schematic of a morphologically patterned substrate, consisting of periodically arranged posts. The post surfaces and the substrate base preferentially attract the A-component of a binary (AB) mixture.}
\end{figure}

From the above discussion, it is clear that fluid mixtures on rough surfaces are experimentally very important. However, to the best of our knowledge, there is no theoretical study of SDSD in these systems. The goal of this paper is to address this shortcoming. We study diffusion-driven SDSD in critical mixtures on morphologically patterned substrates. For simplicity, the patterned substrate consists of periodically arranged posts. Fig.~\ref{fig1} is a
schematic of such a substrate. As in our previous study on chemically patterned substrates, we consider a simple physical pattern to gain theoretical insights. However, many of our results apply to the case of arbitrary surface patterns. Our primary objectives are to study the following issues: \\
1) How does the morphology of domains inside the grooves evolve due to the interaction of SDSD waves originating from the multiple surfaces of the substrate? \\
2) How does the domain morphology depend on the spacing and size of the posts? \\
3) How does the surface pattern affect the kinetics of segregation adjacent to the substrate?

This paper is organized as follows. In Sec.~\ref{sec2}, we describe our modeling and simulation details. The numerical results are presented in Sec. \ref{sec3}. Finally, we summarize our results in Sec.~\ref{sec4}.

\section{Modeling and Simulation Details}
\label{sec2}

We use the PB model \cite{pb92} to study the effect of morphologically patterned substrates on SDSD. We consider an AB mixture in contact with a substrate S. The base of the substrate is located at $z=0$. The height of S is denoted as $H(\vec{\rho})$, where $\vec{\rho}$ are the coordinates parallel to the base. In our simulations, we use a piece-wise flat function $H(\vec{\rho})$. The substrate gives rise to a short-ranged surface potential $V(\vec{\rho},z)$, which acts in a microscopic layer of thickness $a$:
\bea
V(\vec{\rho},z) &=& -h_1(\vec{\rho}), \quad H(\vec{\rho}) \leq z \leq H(\vec{\rho})+a , \nonumber \\
&=& 0, \quad z > H(\vec{\rho})+a .
\eea
We will set $a=0$ subsequently.

In dimensionless units, the free energy functional for an unstable binary mixture in contact with S is \cite{pf97,sp05}
\begin{eqnarray}
\label{free}
\mathcal F\left[\psi (\vec r)\right] &=&{\int d\vec{\rho} \int_{H(\vec{\rho})}^\infty dz \left[ - \frac{\psi^2}{2} + \frac{\psi^4}{4} + \frac{1}{4}(\nabla \psi)^2 \right] } \nonumber \\
&& + {\int d\vec\rho \left[ - \frac{g}{2}\psi (\vec \rho,H)^2 - h_1(\vec{\rho}) \psi (\vec \rho,H) - \gamma\psi (\vec\rho,H) \frac{\partial \psi}{\partial z}\Big|_{z=H} + \frac{\tilde{\gamma}}{2} \left(\vec{\nabla}_\rho
\psi (\vec \rho,H) \right)^2 \right]} \nonumber \\
& \equiv & F_b + F_s.
\end{eqnarray}
In Eq.~(\ref{free}), $\psi(\vec r)$ is the order parameter, which is defined as $\psi(\vec r)=\rho_{\rm A}(\vec r) - \rho_{\rm B}(\vec r)$. Here, $\rho_i (\vec r)$ is the concentration of species $i$ at position $\vec r$. The {\it bulk free energy} $F_b$ has the usual $\psi^4$-form, and $F_s$ is the {\it surface free energy}. In $F_s$, the phenomenological constants $g, \gamma, \tilde{\gamma}$ are related to the bulk correlation length and other system parameters \cite{pf97,sp05}. The one-sided derivative term $\del \psi/\del z |_{z=H}$ appears due to the absence of neighboring atoms for $z<H(\vec{\rho})$. The dependence of surface field strength $h_1$ on $\vec{\rho}$ accounts for any chemical inhomogeneities on the substrate, as considered in our earlier paper \cite{djp20}. In this paper, the substrate is chemically homogeneous with $h_1(\vec{\rho})=h_1$.

We are interested in the kinetics of SDSD, so we consider a time-dependent order parameter field $\psi (\vec{r},t)$. Following PB, this obeys the CHC equation:
\begin{eqnarray}
\label{chceq}
\frac{\del}{\del t} \psi(\vec r, t) = \vec\nabla\cdot\left[\vec\nabla\left\{- \psi + \psi^3 - \frac{1}{2}\nabla^2 \psi \right\} + \vec\theta (\vec r, t)\right] , \quad z>H(\vec{\rho}) .
\end{eqnarray}
In Eq.~(\ref{chceq}), $\vec\theta$ is a vector Gaussian white noise with zero average and dimensionless strength $\epsilon$. The surface is modeled by boundary conditions analogous to those in Ref.~\cite{djp20}:
\begin{eqnarray}
\label{bc1}
\tau_0 \frac{\del}{\del t} \psi(\vec\rho, H, t) &=& h_1(\vec{\rho}) + g\psi(\vec \rho,H,t) + \gamma \frac{\del \psi}{\del z}\Big |_{z=H} + \tilde{\gamma} \nabla^2_\rho \psi(\vec \rho, H, t) , \\
\label{bc2}
0 &=& \left[\frac{\del}{\del z}\left\{ - \psi + \psi^3 - \frac{1}{2}\nabla^2\psi \right\} +
\theta_z\right]_{z=H} .
\end{eqnarray}
Eq.~(\ref{bc1}) describes nonconserved relaxational kinetics (with time-scale $\tau_0$) of the order parameter at the surface. Eq.~(\ref{bc2}) sets the $z$-component of the surface current to zero, as there is no flux across the substrate. The quantities $h_1(\vec{\rho}), g,\gamma$ determine the equilibrium phase diagram of the system~\cite{pf97,sp05}.

Before proceeding, it is important to discuss the relevance of the phenomenological CHC model to the sophisticated applications mentioned earlier, e.g., microfluidics. In this context, we make the following remarks: \\
1) We have not included hydrodynamic effects in our model, though this can be easily done \cite{ht01}. Therefore, the results reported here correspond to diffusion-driven phase separation. However, even for the bulk phase separation of fluids, there is an extended diffusive regime prior to the hydrodynamic regime \cite{pw09}. Thus, our results apply in this regime for SDSD in fluid mixtures. Moreover, the structured substrates considered here disable hydrodynamic transport at larger length scales. Therefore, even at late times, we expect diffusive transport to play an important role for SDSD in fluids. \\
2) The coarse-grained models used here do not apply at molecular length scales. Thus, our results are relevant when the surface structures are at least 1-2 orders of magnitude larger than the molecular scales. \\
3) We know that {\it minimal models} can often capture physics even at length-scales and time-scales at which they are not strictly applicable. Therefore, we expect that the universal behaviors studied here (e.g., power-law exponents, dynamical scaling) would be of greater applicability than the restrictions 1) and 2) above would suggest.

In Sec.~\ref{sec3}, we will present comprehensive numerical results obtained from a Langevin simulation of the above model. Before proceeding, we provide the details of our simulations, which were performed in $d=2$ (system size $L_x \times L_z = 1024 \times 256$), and $d=3$ (system size $L_x \times L_y \times L_z = 256^2 \times 160$). The morphologically patterned substrate consisted of rectangular posts of size $M_x \times M_z$ in $d=2$, and $M_x \times M_y \times M_z$ in $d=3$ (see Fig.~\ref{fig1}). We set $M_x=M_y=M_z=M$, and the linear spacing of the posts was also taken as $M$ in the $x,y$-directions. In our dimensionless rescaling, the length scale is the bulk correlation length $\xi_b$ \cite{djp20}. The pattern sizes are measured in units of $\xi_b$ (e.g., $M=8,16,32$). Thus, the substrate is modulated on a macroscopic length scale. This is necessary for consistency with our coarse-grained CHC model. The post surfaces and the base attracted A with a surface field $h_1 = 1.0$. The other parameter values were $g=-0.4$, $\gamma=0.4$, which corresponds to a CW equilibrium for a flat surface. The lateral diffusion constant was set as $\tilde{\gamma} = 0.0$ as the order parameter is uniform on the substrate. The noise amplitude $\epsilon$ = 0.041, corresponding to deep quenches with $T \simeq 0.22T_c$ \cite{pf97,sp05}.

The boundary conditions in Eqs.~(\ref{bc1})-(\ref{bc2}) were applied at $z=H(\vec{\rho})$. We applied flat and zero-flux boundary conditions (mimicking a semi-infinite geometry) at $z=L_z$:
\begin{eqnarray}
\label{flat}
&& 0 = \frac{\del\psi}{\del z} \bigg|_{z=L_z} , \\
\label{nof}
&& 0 = \left[\frac{\del}{\del z}\left\{ - \psi + \psi^3 - \frac{1}{2}\nabla^2\psi \right\} + \theta_z\right]_{z=L_z} .
\end{eqnarray}
Periodic boundary conditions were applied in the $x$- and $y$-directions.

We used the Euler-discretization technique to solve Eqs.~(\ref{chceq})-(\ref{nof}). The discretization mesh sizes were $\Delta x = 1.0$ and $\Delta t = 0.03$, which yield a stable numerical scheme. The initial condition for the $\psi$-field was taken to be a small-amplitude random fluctuation, $\psi(\vec{r},0) = 0 + \phi$, where $\phi$ was uniformly distributed in $[-0.01,0.01]$. The average value $\psi_0=0$ corresponds to a critical composition, consisting of 50\% A and 50 \% B. This initial condition mimicked the disordered state at the time of the quench below $T_c$ ($t=0$). In Sec.~\ref{sec3}, we will also present statistical data for the $d=3$ case. This was obtained as an average over 25 independent runs with different initial conditions and noise realizations.

\section{Detailed Numerical Results}
\label{sec3}

\begin{figure}[t]
\centering
\includegraphics*[width=0.70\textwidth]{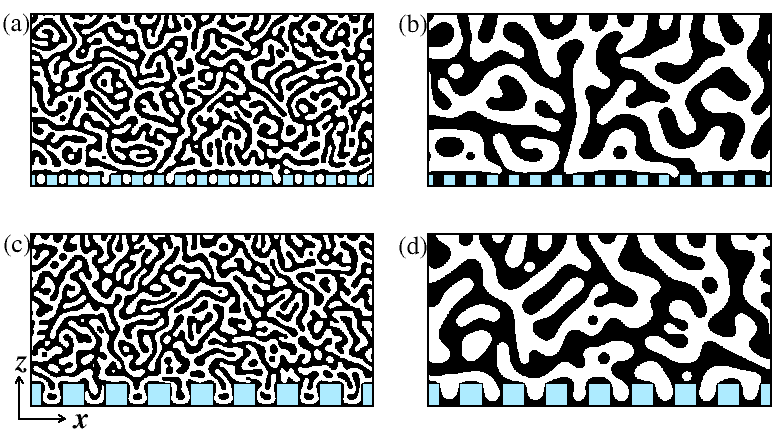}
\caption{\label{fig2} (color online) Snapshots of the order parameter field $\psi(\vec r, t)$ in $d=2$ for a critical mixture. The snapshots correspond to (a) $M=16, t=300$, (b) $M=16, t=5400$, (c) $M=32, t=300$, (d) $M=32, t=5400$. The A-rich regions ($\psi>0$) are marked in black, while the B-rich regions ($\psi<0$) are unmarked. The posts at the bottom ($z=0$) are marked blue. The post sizes are $M \times M$, and they are spaced $M$ apart. The system size is $L_x \times L_z$ with $L_x=1024, L_z=256$. For clarity, we only show the $\psi$-field for a region of size $512 \times 256$.}
\end{figure}

Let us first discuss some representative results for the $d=2$ case. This will set the stage for our discussion of the $d=3$ results. In Fig.~\ref{fig2}, we present evolution snapshots for $d=2$ systems with post sizes $M=16$
(upper frames) and $M=32$ (lower frames). The posts are located in the region $z \in [0,M]$, and are wetted by A. Let us focus first on the evolution for $M=16$ in Figs.~\ref{fig2}(a)-(b) The posts/grooves rapidly develop a layer rich in A followed by a depletion layer -- all surfaces give rise to SDSD waves which propagate away from the surface. There is an interesting linear stability analysis by Fischer et al. \cite{fmd97,fmd98}, which identifies the dominant modes in the early stages of SDSD on a flat substrate. It would be useful to perform a similar analysis in the present context. At early times ($t=300$ in Fig.~\ref{fig2}(a)), the depletion layers of these waves merge so that the grooves are B-rich. At later times ($t=5400$ in Fig.~\ref{fig2}(b)), the grooves are filled by A and the wetting layer blankets the physical pattern. For times greater than the ``filling time'' $t_f$, the evolution is analogous to that for a flat substrate \cite{pb92,pb01}. As usual, we see bulk phase separation for ``large'' values of $z$ -- we will see shortly how the depth profiles of the order parameter are used to identify the bulk. A similar scenario applies for $M=32$ in Figs.~\ref{fig2}(c)-(d) with the following differences: \\
(a) At early times ($t=300$), we see droplets of the A-phase in the grooves. These arise from the union of secondary wetting layers in the SDSD waves originating from the post/groove surfaces. \\
(b) As expected, the filling time $t_f$ increases with $M$. Thus, the grooves are still unfilled at $t=5400$ for $M=32$ -- see Fig.~\ref{fig2}(d). The wetting-layer growth obeys $R_1(t) \sim t^{1/3}$ \cite{pb01}, suggesting that $t_f \sim M^3$.

\begin{figure}[t]
\centering
\includegraphics*[width=0.70\textwidth]{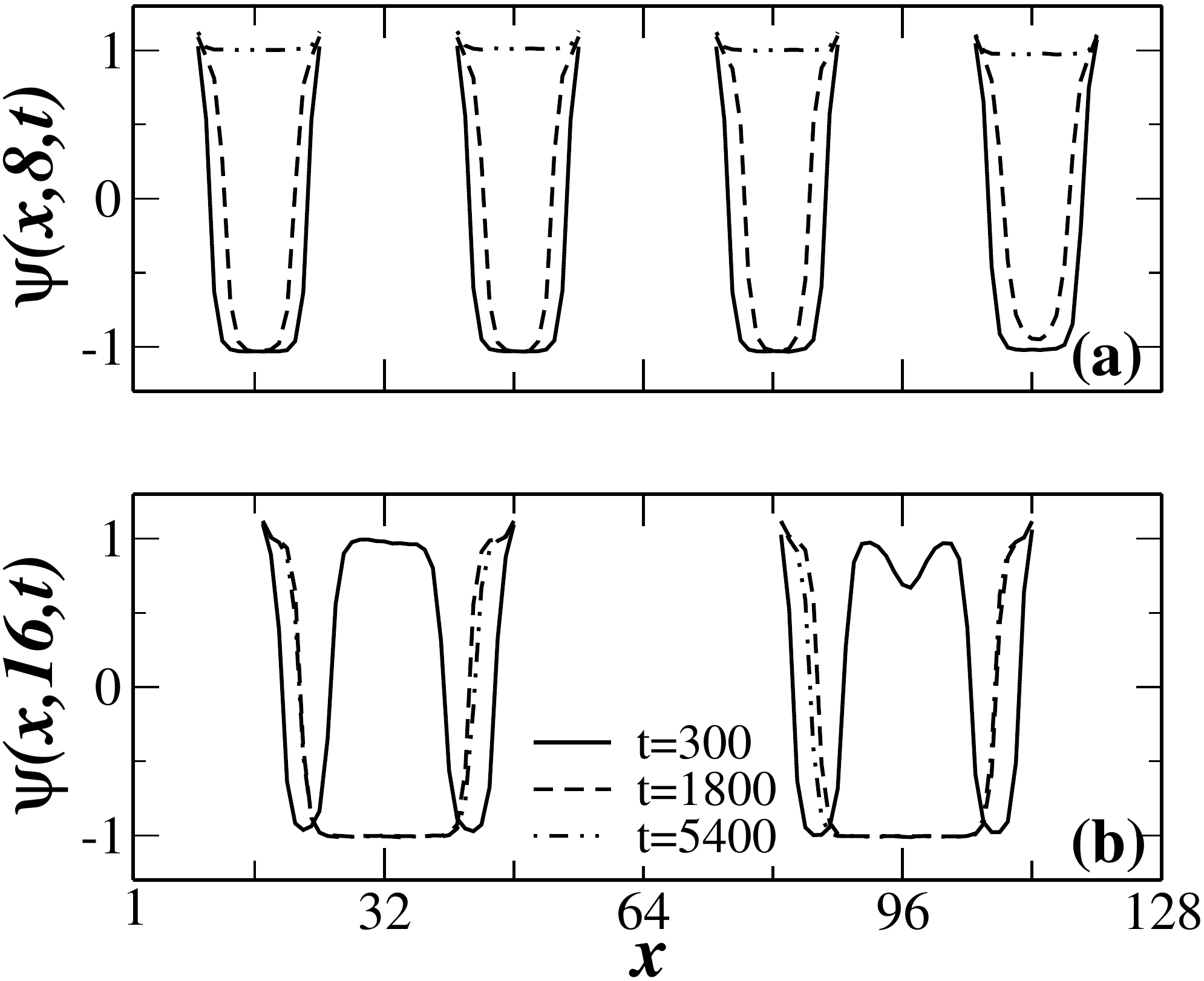}
\caption{\label{fig3} Order parameter profiles for the snapshots in Fig.~\ref{fig2} at (a) $z=8$ for $M=16$, (b) $z=16$ for $M=32$. We plot $\psi (x,z,t)$ vs. $x$ at $t=300,1800$ and $5400$, as indicated.}
\end{figure}

In Fig.~\ref{fig3}, we show the order parameter profiles in the snapshots of Fig.~\ref{fig2} at $z=M/2$. The upper
frame [Fig.~\ref{fig3}(a)] plots $\psi (x,M/2,t)$ vs. $x$ for $M=16$ at $t=300,1800,5400$. The lower frame [Fig.~\ref{fig3}(b)] is the analogous plot for $M=32$. This figure confirms the observations made above in the context of interference of the SDSD waves.

\begin{figure}[t]
\centering
\includegraphics*[width=0.60\textwidth]{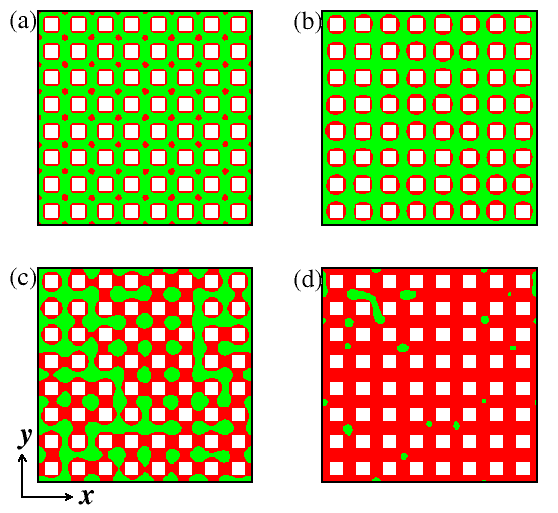}
\caption{\label{fig4} (color online) Snapshots of the $\psi$-field in $d=3$ at (a) $t=90$, (b) $t=900$, (c) $t=1800$, and (d) $t=3150$. The system size is $256^2 \times 160$, and the post size is $16^3$ with a spacing of 16. We show snapshots at $z=8$ (in the grooves). The A-rich regions ($\psi > 0$) are marked red, and the B-rich regions ($\psi < 0$ are marked green. The posts are unmarked.}
\end{figure}

Let us next present detailed results for the $d=3$ case. Our results correspond to post sizes $M=16$ with a spacing $M=16$, unless mentioned otherwise. In Fig.~\ref{fig4}, we show snapshots of the $\psi$-field inside the
grooves at $z=M/2$. The times in Fig.~\ref{fig4} are all less than $t_f$. For $t=90$ [Fig.~\ref{fig4}(a)], the SDSD waves from the post/groove surfaces show secondary wetting layers, which interfere to form a regular array of A-rich droplets. Clearly, an appropriate lay-out of posts on the substrate could be used to tailor any desired mesoscale structure. In the present case, the array of droplets is transient, but it can be frozen in several ways. The secondary wetting layers are gone by $t=900$ [Fig.~\ref{fig4}(b)], leaving only the wetting and depletion layers. The filling process has begun by $t=1800$ [Fig.~\ref{fig4}(c)], and is almost complete by $t=3150$ [Fig.~\ref{fig4}(d)].

\begin{figure}[t]
\centering
\includegraphics*[width=0.80\textwidth]{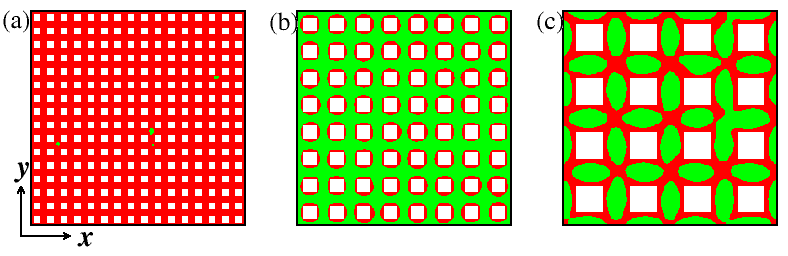}
\caption{\label{fig5} (color online) Snapshots of the $\psi$-field for a critical mixture at $t=900$. The color coding is the same as Fig.~\ref{fig4}. We show snapshots at $z=M/2$ for (a) $M=8$, (b) $M=16$, and (c) $M=32$.}
\end{figure}

In Fig.~\ref{fig5}, we study the effect of the post size on the morphology inside the grooves. We plot the
$\psi$-field at $t=900$ and $z=M/2$ for $M=8,16,32$ in Figs.~\ref{fig5}(a)-(c), respectively. As discussed earlier, the filling is faster for smaller $M$. The snapshot for $M=8$ already shows that the grooves are filled with A up to $z=M/2$. For $M=16$, the depletion layers of the SDSD waves have merged to form a B-rich morphology with A-rich wetting layers on the post surfaces. For $M=32$, the interference of SDSD waves gives rise to a complex and beautiful morphology. The difference in the morphologies for $M=16,32$ is a result of (a) the distance between the posts, which determines the time-scale of interference of the SDSD waves; and (b) the height of the posts, which determines the filling time. Of course, the grooves in the $M=32$ system will only be filled on much longer time-scales ($t_f \sim M^3$). It is difficult to access these time-scales with the Euler discretization scheme used here as this requires a small value of the time increment. However, a more efficient implicit discretization scheme \cite{kem01} would permit simulations up to much longer times.

\begin{figure}[t]
\centering
\includegraphics*[width=0.50\textwidth]{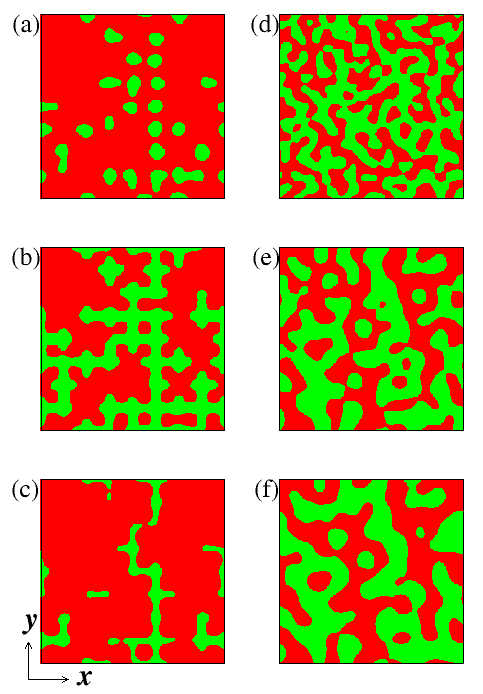}
\caption{\label{fig6} (color online) Snapshots of the $\psi$-field in the ($x,y$)-plane for a critical binary mixture. The color coding is the same as Fig.~\ref{fig4}. The left-hand frames correspond to $z=17$ at (a) $t=900$, (b) $t=3600$, and (c) $t=6300$. The right-hand frames correspond to $z=120$ (bulk) at (d) $t=900$, (e) $t=3600$, and (f) $t=6300$.}
\end{figure}

We next focus on the evolution morphology above the grooves for $M=16$. The left panels of Fig.~\ref{fig6} show
the evolution at $z=17$ (just above the grooves). The drainage of A into the grooves results in a complex evolution at $z=17$, with the average value of $\psi$ changing substantially with time. (This will be confirmed shortly.) However, the imprint of the underlying pattern is seen in the snapshots for $t < t_f$. The picture at $t=6300$ [Fig.~\ref{fig6}(c)] corresponds to the grooves being almost full. For $t > t_f \simeq 7000$, the $z=17$ layer is absorbed into the growing wetting layer. By this time, the surface pattern has been blanketed by the wetting layer. As stated earlier, the subsequent evolution is analogous to that for the flat substrate. For purposes of reference, the right panels of Fig.~\ref{fig6} show the evolution at $z=120$, corresponding to the bulk. These snapshots show the typical bicontinuous morphology of spinodal decomposition in a critical mixture.

\begin{figure}[t]
\centering
\includegraphics*[width=0.50\textwidth]{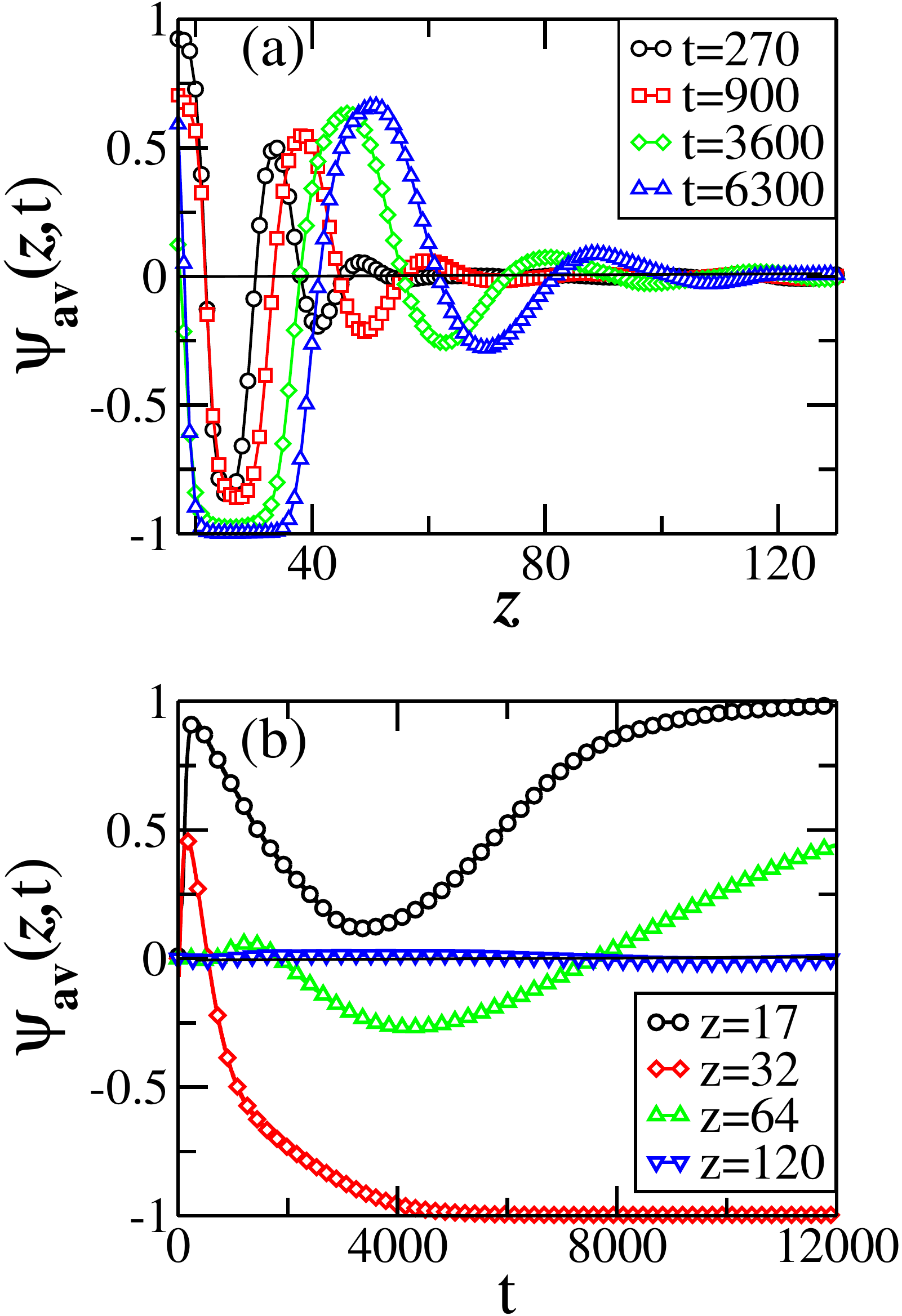}
\caption{\label{fig7}(color online) (a) Laterally averaged profiles, $\psi_{\rm av}(z,t)$ vs. $z$. We show data for 4 different times, as specified. (b) Plot of $\psi_{\rm av}(z,t)$ vs. $t$ for $z=17,32,64,120$ (bulk).}
\end{figure}

How do we distinguish between the surface region and the bulk? In Fig.~\ref{fig7}(a), we plot the 
laterally-averaged order parameter profiles $\psi_{\rm av}(z,t)$ vs. $z$ \cite{jnk91,pb92} at 4 different times. These are obtained from a typical evolution in Fig.~\ref{fig6} by averaging $\psi (x,y,z,t)$ in the $x$- and $y$-directions. We further average the data over 25 independent runs. The ``surface region'' shows a systematic profile, whereas the ``bulk region'' has $\psi_{\rm av}$ fluctuating around 0. Clearly, the surface region propagates into the bulk with time. These depth profiles show a {\it fast mode} and a {\it slow mode}. The fast mode corresponds to the relaxational dynamics of $\psi$ due to the boundary conditions. The substrate rapidly becomes enriched in A and $\psi_{\rm av}(z=17,t) \simeq 1.0$ at $t=270$. Then, the slow mode driven by bulk phase separation becomes operational. This destroys the early-time layered structure and $\psi_{\rm av}(z=17,t) \simeq 0.2$ at $t=3600$. The system then develops a stable SDSD profile consisting of a wetting layer followed by a depletion layer, which propagates into the bulk. The secondary wetting layer seen at $t=6300$ is ultimately destroyed by the bulk phase separation. The surface profile extends up to $z \simeq 60,70,90,100$ at times $t=270,900,3600,6300$, respectively. In Fig.~\ref{fig7}(b), we plot $\psi_{\rm av}(z,t)$ vs. $t$ for 4 different values of $z$. This shows how the off-criticality in different layers evolves with time. It is non-monotonic for $z=17,20$ because of the presence of the fast and slow modes seen in Fig.~\ref{fig7}(a). For $z=120$, there is no systematic time-dependence ($\psi_{\rm av} \simeq 0.0$) over the window of our simulation. Hence, we consider $z=120$ as lying in the bulk region.

\begin{figure}
\centering
\includegraphics*[width=0.40\textwidth]{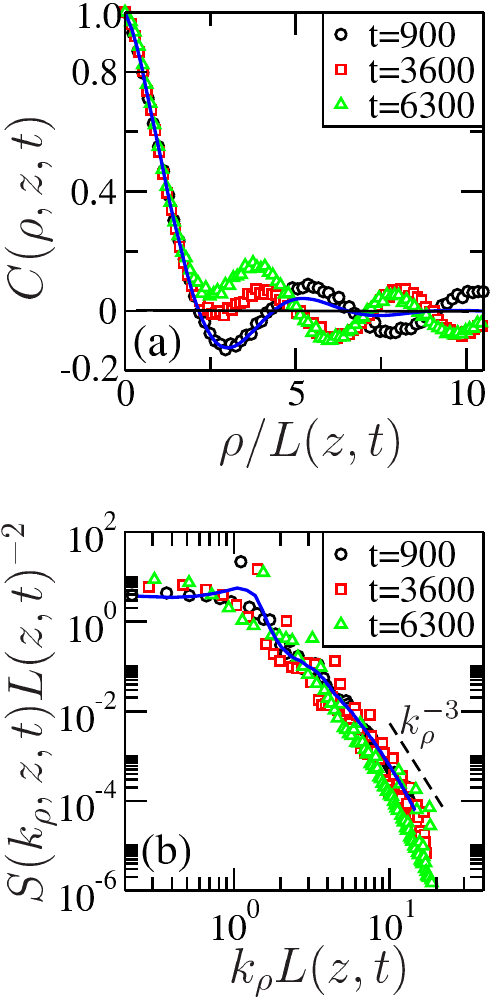}
\caption{\label{fig8} (color online) (a) Scaling plot of layer-wise correlation functions, $C(\rho,z,t)$ vs. $\rho/L(z,t)$ for $z=17$ and 3 different times, as mentioned. The length scale $L(z,t)$ is the distance over which $C(\rho,z,t)$ falls to half its maximum value [$C(\rho=0,z,t)=1$]. The solid line denotes the $z=120$ data at $t=5400$, which corresponds to the bulk. (b) Scaling plot of layer-wise structure factors, $S(k_\rho,z,t) L(z,t)^{-2}$ vs. $k_\rho L(z,t)$, for the data in (a). The dashed line labeled $k_\rho^{-3}$ denotes the Porod law.}
\end{figure}

Let us quantitatively characterize the evolution in layers adjacent to the posts. This evolution was shown pictorially for $z=17$ in Fig.~\ref{fig6}. For this purpose, we calculate the {\it layer-wise correlation function}:
\begin{eqnarray}
\label{corr}
C(\vec{\rho}, z, t) = \frac{1}{L_x\times L_y} \int d\vec R\left[\langle \psi(\vec R, z, t) \psi(\vec R+\vec\rho, z, t)\rangle - \langle\psi(\vec R, z, t)\rangle \langle\psi(\vec R+\vec\rho, z, t)\rangle\right] .
\end{eqnarray}
As the morphology is isotropic in the $(x,y)$-plane, we spherically average $C(\vec{\rho},z,t)$ to obtain $C(\rho,z,t)$. Fig.~\ref{fig8}(a) is a scaling plot, where we superpose data for $C(\rho,z,t)$ vs. $\rho/L(z,t)$
from 3 different times $< t_f \simeq 7000$. The length scale $L(z,t)$ is defined as the distance over which $C(\rho,z,t)$ decays to half its maximum value. We see that the data does not scale because the off-criticality changes continuously over this time-window -- see Fig.~\ref{fig7}(b). The scaling function is known to depend on the off-criticality \cite{pw09}. In Fig.~\ref{fig8}(a), we also plot the bulk scaling function, obtained from the 
$z=120$ data. The {\it layer-wise structure factor} $S(\vec{k}_\rho,z,t)$ is defined as the Fourier transform of $C(\vec{\rho},z,t)$ at wave vector $\vec k_\rho$:
\begin{eqnarray}
\label{struc}
S(\vec{k}_\rho,z,t) = \int d\vec\rho~e^{i\vec{k_\rho} \cdot \vec\rho}~C\left(\vec\rho, z, t\right) .
\end{eqnarray}
We spherically average $S(\vec{k}_\rho,z,t)$ in the $\vec{k}_\rho$-plane to obtain $S(k_\rho,z,t)$. The dynamical scaling form of $S(k_\rho,z,t)$ is the appropriate generalization of Eq.~(\ref{scale2}). 
\beq
S(k_\rho,z,t) = L(z,t)^d f \left[k_\rho L(z,t) \right] .
\eeq
In Fig.~\ref{fig8}(b), we plot scaled data for $L(z,t)^{-2} S(k_\rho,z,t)$ vs. $k_\rho L(z,t)$ for the same times as in Fig.~\ref{fig8}(a). The bulk data is shown as a solid line. In the limit of large $k_\rho$, the structure factor data is consistent with the well-known \textit{Porod's law} \cite{gp82,op88}: $S(k_\rho,z,t) \sim k_\rho^{-3}$. This is a consequence of scattering from sharp interfaces, which are always present in the domain morphology (see Fig.~\ref{fig6}).

\begin{figure}[t]
\centering
\includegraphics*[width=0.50\textwidth]{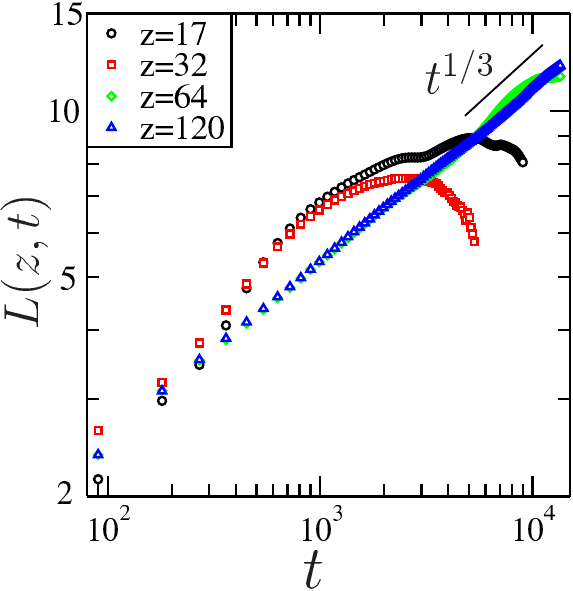}
\caption{\label{fig9} (color online) Log-log plot of layer-wise domain scale, $L(z,t)$ vs. $t$. We show data for different values of $z$, as specified. The line labeled $t^{1/3}$ denotes the LS growth law.}
\end{figure}

Finally, in Fig.~\ref{fig9}, we show the time-dependence of the layer-wise domain scale: $L(z,t)$ vs. $t$ for
$z=17,32,64,120$. The bulk data ($z=120$) is consistent with the LS growth law, $L(t) \sim t^{1/3}$, as expected. Similarly, the $z=64$ data is consistent with the LS growth law up to $t \simeq 8000$. The deviation seen for $t > 8000$ is a result of the $z=64$ region entering the surface region -- see Fig.~\ref{fig7}. The layers with $z=17,32$ are strongly off-critical for almost the entire duration of our simulation (see Fig.~\ref{fig7}). Therefore, we do not observe a systematic behavior for the ``growth'' of fluctuations in this off-critical background.

\section{Summary and Discussion}
\label{sec4}

Let us conclude this paper with a summary and discussion of our results in this two-part exposition. We have used the Puri-Binder (PB) model \cite{pb92} to numerically study {\it surface-directed spinodal decomposition} (SDSD) on patterned substrates. Such substrates may arise naturally, or they may be structured with specific functionalities in mind. There has been huge experimental interest in SDSD on patterned substrates, due in large part to their multiple technological applications. However, our theoretical understanding of this problem is very poor. In the case of chemically heterogeneous surfaces, there have been some preliminary theoretical studies of the emergent morphologies. On the other hand, for morphologically patterned substrates, there appear to be no such studies to the best of our knowledge.

In the present exposition, we have tackled the ambitious task of providing a quantitative theoretical basis for SDSD on patterned substrates. In our first paper \cite{djp20}, we discussed surfaces which were chemically heterogeneous but physically flat. In this paper, we have focused on surfaces which are chemically homogeneous but physically non-uniform. In both papers, we have considered substrates with simple representative patterns. However, many of our results apply to the case of arbitrary surface patterns.

A summary of the results in our first paper has already been presented therein. Our salient results in this second paper are as follows: \\
(a) In the grooves, the interference of SDSD waves can be used to tailor specific structures in the phase-separating mixture. A simple checkerboard array, as considered here, already yields a range of fascinating morphologies on adjusting the size and spacing of the posts. These mesoscale structures can be frozen by a variety of methods, e.g., a photo-induced chemical reaction A $\rightleftharpoons$ B \cite{pf94}, introduction of quenched disorder, etc. \\
(b) Once the grooves are filled (i.e., after the filling time $t_f$), SDSD on a morphologically patterned substrate is analogous to that on a flat surface. \\
(c) There is a complex pattern dynamics just above the grooves, due to the propagation of the wetting layer into the bulk. This can be characterized by layer-wise correlation functions, structure factors, growth laws, etc. As expected, the bulk of the system exhibits dynamical scaling and Lifshitz-Slyozov (LS) growth. However, the boundary which demarcates the {\it bulk region} from the {\it surface region} is time-dependent. We were surprised to see how deep the surface region can extend at late times. We emphasize that this is not due to the surface-molecule interactions which have a microscopic range. Rather, it is because of the establishment and propagation of SDSD waves. \\
The above results (a)-(c) apply for arbitrary surface patterns with appropriate modifications. Moreover, we also expect scaling laws to be robust. For example, the filling time for a generic surface pattern of height $\sim H$ should scale as $t_f \sim H^3$.

This is a fascinating subject which clearly requires much more theoretical investigation. For example, we have focused on segregation via diffusive transport, as in solid mixtures. However, many experiments involve fluid mixtures, where velocity fields can alter the kinetics of SDSD \cite{bpl01,jpd12}. The role of hydrodynamics in SDSD on patterned substrates is an open problem of great interest. Further, in the present study, we have studied critical mixtures which consist of 50 \% A and 50 \% B. The introduction of {\it off-criticality} and {\it nucleation} would significantly affect the emergent morphologies \cite{pb01}. The above are just two classes of problems, out of many possible future directions. It is our hope that the present studies will generate fresh experimental and theoretical interest in the classical problem of SDSD. \\
\ \\
\noindent{\bf Acknowledgments:} PD acknowledges financial support from the Council of Scientific and Industrial Research, India.

\end{document}